\documentstyle[11pt,aaspp4,flushrt]{article}
\begin{document}
\newcommand{\beq}{\begin{equation}}
\newcommand{\eeq}{\end{equation}}

\title{
Analysis of Time Delays in the Gravitational Lens 
PG1115+080}

\author{Rennan Bar-Kana\footnote{email: barkana@arcturus.mit.edu}}
\affil{Physics Dept.\ Rm.\ 6-218M, MIT, 77 Massachusetts Ave,
Cambridge, MA 02139 USA}

\begin{abstract}
We present a new method for determining time delays among the light
curves of various images in a gravitational lens. The method
is based on constructing a simple model for the source variation
and forming a $\chi^2$ measure of the agreement of this same 
variation with all of the lightcurves. While inspired by
Press et al.\ (1992a, b) our approach is different since we do 
not assume a Gaussian process for the source variation.
Our method has a number of desirable properties: first, it yields 
an approximate reconstruction of the
source variation and of other parameters such as relative
time delays; second, it easily incorporates different assumptions 
about the relations among the light curves and about the data 
measurement errors; finally, it can be applied to more than two 
light curves by addition of $\chi^2$. We apply this method to
the light curves of the quadruple gravitational lens
PG1115+080 measured by Schechter et al.\ (1997). Unlike 
Schechter et al.\ we include correlated measurement errors
in the analysis, as well as the possibility that microlensing
may cause different images to vary by different factors in flux.
We find a value of $25.0^{+3.3}_{-3.8}$ days ($95\%$ confidence) for the 
delay between components B and C (close to the $24$ day value of Schechter 
et al., and so leading to a similar value of the Hubble constant for
a given lens model). However, the ratio $t_{AC}/t_{BA}$ of the two 
intermediate delays is poorly determined at 
$1.13^{+.18}_{-.17}$ ($68\%$ confidence), close to the value
predicted by lens models ($\sim 1.4$) unlike the Schechter et al.\ 
value ($\sim 0.7$). The variation ratios of C with respect to A and
of A with respect to B are both different
from 1, $1.39^{+.16}_{-.20}$ and $.79^{+.10}_{-.12}$ ($95\%$ confidence),
respectively. This is an indication of a microlensing gradient, and this 
type of microlensing may allow us to conclude that the size of the quasar 
optical emission region is about $1000$ AU.
\end{abstract}


\vspace{.1in}

\section{Introduction}

Even before the discovery of the first gravitational lens 
0957+561 (\cite{0957dis}) it was recognized that 
measurements of the time delay between images can be used to
determine the Hubble constant (\cite{refsdal64}, 1966).
Despite a history of systematic difficulties, recent measurements combined
with an analysis of lens models (\cite{gn}) indicate 
that a robust measurement of the Hubble constant ($H_0$) with an accuracy 
comparable to that of more conventional techniques may be within reach.
The measurements include a precise optical time delay
(Kundi\'{c} et al.\ 1997) which is consistent with the latest 
results from radio monitoring (Haarsma et al.\ 1997), and a measurement
of the velocity dispersion of the lens galaxy (Falco et 
al.\ 1997).

Since each lens can potentially yield an independent, single
step $H_0$ measurement, obtaining values in
a number of lenses with a variety of morphologies and constraints
on models may help eliminate systematic errors. Each lens
requires a well-constrained lens model, lens and source redshifts,
and a measurement
of the time delays among the images. Observational constraints 
on the lens model may include precise image positions and flux
ratios as well as data on the objects responsible for the lensing.
Resolved structure in images as revealed by radio interferometry
provides many constraints, since it is essentially the same as
observing multiple sources with the same lens. Whether lensing
involves a galaxy, group, or cluster, the position, velocity
dispersion, and other observational probes of the mass distribution
of the lensing objects all yield constraints on lens models.
Measuring time delays requires observing time variations in 
the image intensities with sufficient accuracy and time resolution. 
As the number of lenses being carefully monitored increases,
more time delays are being determined, such as the
preliminary measurement in PKS 1830-211 (\cite{pks}). 
Flux measurements are not the only possibility,
as shown by the promising measurements of variations in 
polarization fraction in the images of B 0218+357 (\cite{iau137}).

The quadruply imaged quasar PG1115+080 was the second lens
to be discovered (\cite{weymann}). It is radio quiet, but
optical Hubble Space Telescope images (\cite{hst1115}) were
recently analyzed by Schechter et al.\ (1997, hereafter SCH) and by
Keeton \& Kochanek (1997). They found that lens models which include 
the effect of the lens galaxy and that of the nearby group of galaxies 
discovered by Young et al.\ (1981) can fit the image positions 
well. They however still found great freedom in the $H_0$ values 
predicted by these lens models for a given time delay. 

In four-image configurations where the images lie
at roughly the same distance from the lens, there
is a well known degeneracy between the radial profile of
the lens mass and the inferred $H_0$ (\cite{deg1}; \cite{deg2}).
In this situation the relative image positions do determine
the total enclosed mass within the ring of images, but they
are not very sensitive to the radial profile of the mass. Changing
the radial profile affects the convergence at the images,
and this changes the inferred $H_0$ value in a similar
way to the effect of a constant mass sheet (\cite{falco1};
\cite{narayan}). In PG1115+080 future observations of the
lensing galaxy light profile and, ultimately, a direct
measurement of its central velocity dispersion may constrain
or eliminate this degeneracy.

Recently SCH measured light curves for
the A1, A2, B and C images of PG1115+080, and used them to
determine multiple delays. The bright A1 and A2 images 
are close together and should have a very small time delay
($\sim$ a few hours), so they were combined into a single
A=A1+A2 curve. SCH used the Press et al.\ (1992a, b) 
method and found that C leads A and A leads B,
with $t_{BC}=23.7\pm3.4$ days and $t_{AC}=9.4\pm3.4$ days 
which yields a ratio $r_{ABC}\equiv t_{AC}/t_{BA}$ of 
$0.7\pm0.3$. It is useful to express the two independent
quantities as $t_{BC}$ and $r_{ABC}$, since $t_{BC}$ can
be taken to fix the Hubble constant for a given lens model 
while $r_{ABC}$ is independent of overall distance, and can be 
compared directly with the ratio predicted by lens models.
The models mentioned above are consistent in predicting
$r_{ABC}=1.4$ to within about $0.1$, and SCH
noted the $2\sigma$ discrepancy with their fitted value.
In their analysis SCH assumed that the measurement errors in 
the light curves are uncorrelated, and also that the fractional
flux variations are the same for each component.

In this paper we present a more detailed analysis of the
light curves in PG1115+080. We first present a new method 
based on $\chi^2$ minimization, which has many of the
advantages of Press et al.\ (1992a, b), but is simpler 
and allows for a more 
conservative assessment of errors in the reconstructed
parameters. We then apply this method to PG1115+080, and
include correlated measurement errors in the analysis.
In addition to relative time delays, we also allow for 
different factors of variation in flux, which may
arise from microlensing of the images.
In \S 2 we present our $\chi^2$ method, and discuss its distinct
features and free parameters. We then discuss the physical 
meaning of the various parameters that the method can account for
and attempt to extract from data. In \S 3 we apply our method
to fitting the PG1115+080 light curves, singly, in pairs, and
all together, and discuss the results and implications. Finally
in \S 4 we summarize our results and point out some of the significant
returns possible from further monitoring.

\section{Method and physical parameters}

Suppose we have a light curve of $N$ points, $(t_i,v_i)$
 $i=1,\ldots,N$, where $t_i$ is time (in days) and $v_i$ is the 
intensity (in magnitudes) at time $t_i$, and we construct a model 
for $v(t)$ with $M$
adjustable parameters $a_j$, $j=1,\ldots,M$. Then if the
$v_i$ measurements are independent, with normally distributed
errors of standard deviation $\sigma_i$ for each $i$, the
$\chi^2$ measure of goodness-of-fit is 
\beq
\chi^2=\sum_{i=1}^N \left[\frac{v_i-v(t_i;a_1\ldots a_M)}
{\sigma_i}\right]^2\ .
\eeq
When the measurement errors are correlated, with a covariance
matrix ${\bf Q}$, then 
\beq
\chi^2=\sum_{i,j=1}^N W_{ij}\ [v_i-v(t_i;a_1\ldots a_M)]\ 
[v_j-v(t_j;a_1\ldots a_M)]\ ,
\eeq
where the weight matrix ${\bf W}$ is the matrix inverse
of ${\bf Q}$.

We wish our model for $v(t)$ to be an arbitrary continuous 
curve. Although quasar variations may be stochastic,
the variations typically decrease in amplitude at smaller
and smaller time scales, and a continuous curve can
describe the large time-scale variations well. Of course,
we cannot take an {\it arbitrary}\/ curve, since we must
describe the curve with a small number of parameters.
We therefore take $L$ points $(t_k,v_k)$ with the $t_k$
just covering the range of the data and equally spaced.
The $v_k$ are then the parameters, and the model for
$v(t)$ is a curve going through all the $(t_k,v_k)$ with
interpolation used for other values of $t$. Note that
we interpolate between the $v_k$ in the model, only as
a way of parameterizing a general curve $v(t)$ in terms
of the $L$ parameters; we do
not interpolate between the noisy data points directly.
When we fit multiple light curves, a single such model is 
used for all of them, or equivalently the light curves
are all combined into a single light curve which is
then compared to the model. Additional parameters
can be added to account for relative shifts of an entire
light curve relative to the others, before comparing to
the model. A relative magnification
corresponds (in magnitudes) to adding a constant to
all the measured $v_i$ of an image. A relative time delay
corresponds to adding a constant to all the measured
$t_i$ of an image. For given time delays, all the other
parameters are linear (i.e.\ the $\chi^2$ is a quadratic
form in these parameters) if we use only interpolation
which is linear in the $v_k$ (but arbitrarily non-linear
in the $t_k$). Minimizing the $\chi^2$ then corresponds
to solving a linear system, and we solve it using
Cholesky decomposition (e.g.\ \cite{nr}, \S 2.9).

Like Press et al.\ (1992a, b), our method compares a model
for the source variation to the light curves with a 
$\chi^2$ measure, and thus both methods assume that the
statistics of the measurement errors are Gaussian. However,
the two methods construct the model of the source variation
very differently. In Press et al.\ the source variation
is assumed to be a stationary Gaussian process, with a
correlation function which is extracted from the light
curves themselves. In that method there is some freedom to
vary the assumed correlation function in order to check for 
robustness. In our method we assume no particular form for
the correlations among source fluxes at different source
times. By taking a continuous curve we are effectively
smoothing the actual source variation over some small
time scale, but the curve is otherwise arbitrary.
 Although the two methods use different
assumptions, they may not necessarily yield very
different results, since the intrinsic variability of quasars
appears to decrease at small time scales. Qualitatively,
then, both methods are dominated by requiring smoothness
for nearby points when all the light curves are combined.
We also note that the correlations which we {\it are}\/ including
are in the observational errors of different images 
observed at a single time. Such correlations were not
accounted for by SCH, but they in principle can also be
included in the Press et al.\ method.

Our model is very flexible. There is freedom to choose 
both the number of interpolation points L and the
interpolation method. We use several methods of 
interpolation. Given some time $t$ between $(t_k,v_k)$ 
and $(t_{k+1},v_{k+1})$, in linear interpolation
we take for $v(t)$ the linearly interpolated
value between $v_k$ and $v_{k+1}$. In cubic interpolation
we interpolate using the cubic polynomial through the four 
points at $k-1$, $k$, $k+1$ and $k+2$ (with linear interpolation
used for $k=1$ or $k=L-1$). In cubic spline interpolation
(e.g.\ \cite{nr}, \S 3.3), we draw a cubic polynomial between
every pair of points $k$ and $k+1$ so that the entire curve is 
continuous in the second derivative. We find in \S 3 below that 
changing the interpolation method has a rather minor effect. 
Note that with any of these interpolation methods there
is some inter-dependence among all of the $v_k$, unless
there is some $k_0$ with no data points between $t_{k_0}$
and $t_{k_0+1}$. In this case the parameters $v_k$ for
$k>k_0$ would decouple from those with $k<k_0$, at least with
linear interpolation. There are $37$ points in each light curve 
in the PG1115+080 data, and the final data point
follows a gap of 38 days. Hereafter we drop this last point, 
and then we are never in a situation where such a 
$k_0$ exists (The above argument suggests that the last point
has only a small effect on the results with our method, and
this is indeed the case).

The number of points in each light curve is thus
$36$, and we use an $L$ of $10$, $20$ and $30$.
Below we find that the differences in the derived parameters
using the various $L$ are fairly large, so this freedom
provides a more conservative and robust estimate of
parameter uncertainties. Using small $L$ corresponds
essentially to a larger-scale smoothing of the data,
though once again the smoothing is not directly of
the raw data. If we use a very small $L$, we are
over-smoothing the signal, and we may expect an unreliable
reconstruction if the actual signal is not smooth enough
on timescales $\leq \Delta T/L$, where $\Delta T$ is
the total length of the data. If on the other hand we use
a large $L$, there are fewer constraints on the $v_k$,
and we may get large oscillations between nearby $v_k$
as the fitting procedure tries to compensate for 
individual measurement errors. Using a range of $L$ thus 
allows an important check on robustness, and in some sense 
checks for systematic errors over and above the statistical
errors on the results for a given $L$. 

When we use a method based on $\chi^2$ minimization for 
finding time delays among a number of curves, we may
worry about being biased towards large time delays. This is
because large time delays imply little overlap among the
light curves, and so smaller $\chi^2$ since there are fewer 
constraints coming from requiring smoothness when the light curves 
are combined into a single curve. In our method, with a fixed $L$, 
this bias is partially offset since larger time delays
imply larger effective smoothing as the total $\Delta T$
for the combined light curves increases. In practice
we find negligible bias in the Monte Carlo tests
described in \S 3 below.

We wish to also account for the effect of microlensing
in the fitting. Microlensing is the additional magnification
of each image due to lensing by stars or dark matter clumps
in the lensing galaxy. In 2237+0305 microlensing on the scale
of months was first detected by Irwin et al.\ (1989) and confirmed 
by Houde \& Racine (1994). Such microlensing has also been 
detected in 0957+561, with indications of possible microlensing
on smaller time scales (\cite{schild}). Microlensing may cause 
an amplification of order $0.6$ magnitudes, which in PG1115+080 is 
expected to vary over a time scale (\cite{micro}) of $\sim 10$ years 
(If there is microlensing on a scale of days or weeks in PG1115+080,
it appears to be small in the data, and we include it in the
noise and not explicitly in the analysis).
However, even if there is no significant variation over
the span of the measured light curves, there may be a
'static' microlensing effect which causes different 
images to vary by different factors in flux, as suggested by
SCH. This results from the fact that when we observe a quasar
varying over time we may effectively be observing two different
sources. If the flux variation results from a uniform brightening
and dimming of the entire quasar emission source, then we are only
observing a single source. However, if some portions of the emission 
region vary differently from other parts, then the region which
produces the mean (i.e.\ time-independent part of the) flux 
is not identical to the region responsible
for the small intensity variations with time. A possible physical 
picture is that the variation is caused by a 
small jet emanating from the core or a sudden localized surge of
mass accretion. 

In such a situation, the overall
magnifications $M_i$ of the mean flux $F$ and $m_i$ of the 
variation $f$ may be unequal for a given image $i$, and their
ratio $m_i/M_i$ may be different for different images. 
In magnitudes, we would then measure (except for an overall
factor of $-5/2$) 
\beq
\log_{10} (M_i F+m_i f)=
\log_{10} (M_i F)+\log_{10} \left(1+\frac{m_i f}{M_i F}
\right) \simeq \log_{10} (M_i F) +\frac{m_i}{M_i}
\frac{f}{F\ \log 10}\ ,
\eeq
where we assume $f/F \ll 1$. The variation in magnitudes
may therefore have different amplitudes in different 
images if the $m_i/M_i$ ratio is not the same for all $i$.

The spatial scale for microlensing is set by the Einstein 
radius of a typical deflector at the lens plane, projected 
onto the source plane. In PG1115+080 this radius is 
\beq
\zeta_E=5 \times 10^{-3}\ \left[\left<M\right>/0.1 
M_{\sun}\right]^{1/2}{\rm\ pc}=1000\ \left[\left<M\right>/0.1 
M_{\sun}\right]^{1/2}{\rm\ AU}\ ,
\label{qsosize}
\eeq
where $\left<M\right>$ is the average stellar mass
(\cite{micro}). In order to have $m_i$ and $M_i$ not be equal, one 
or both of the emission regions must have structure on scales
of order $\zeta_E$ or smaller: if they were both smooth then 
the microlensing would be washed out, and we would have $m_i=M_i$ 
for all $i$. On the other hand, if the two emission regions 
overlapped and their union had a very small extent compared to 
$\zeta_E$, then the magnification due to microlensing would
not vary over this scale, and again we would have $m_i=M_i$.
Thus we reach the interesting conclusion that the indication
of different $m_i/M_i$ for different $i$ (see \S 3 below) implies 
that the size of the quasar optical emission region {\it must be of 
order}\/ $\zeta_E$, say to within a factor of $10$. 

There are, however,
a number of reasons to be cautious about this statement.
The two emission regions may have different sizes, and
in this case we might expect the mean region to contain
the variation region, and then the different $m_i/M_i$ 
would imply the lower bound only on the larger
emission region. More generally, while there appears to
be significant emission from a structure of size $\zeta_E$,
this does not exclude a range of structure also on smaller or
larger length scales. In addition, $\zeta_E$ depends on 
what we assume for $\left<M\right>$, and there could be a large
range of masses contributing. The intensity variation may
involve relativistic motions, and such high velocities can
cause microlensing variations on small time scales that we
might confuse as differences in $m_i/M_i$. 
The stochastic nature of optically-thick microlensing
(e.g.\ \cite{micro}) implies the possibility of a rare 
statistical fluctuation over the relatively short time span 
of the current data. Note also that the 
magnification due to the macrolens itself varies with position, 
but it should only vary over much larger spatial scales.
There could be a different amount of contamination by light
from the lens galaxy in each image, and this would lead to
apparent changes in the $M_i$ with no corresponding changes
in the $m_i$. However, in PG1115+080 this effect seems to
be too small to affect the $m_i/M_i$ ratios significantly.
Finally, there may still be unmodelled systematic errors in 
the measured light curves which could give the appearance 
of unequal $m_i/M_i$ ratios.
With a longer time span of measurements it may be 
possible to eliminate some or all of these uncertainties. 
See also Gould \& Miralda-Escud\'{e} (1997) for an independent,
illuminating discussion of the possible observational consequences
of differential microlensing of accretion disks in quasars.
We note that the size given in equation (\ref{qsosize}) is
roughly consistent with the upper limit from the interpretation of
microlensing in 2237+0305 (Wambsganss, Paczynski \& Schneider 1990).
It is somewhat smaller than an estimate by Schild (1996) based on 
variations in 0957+561.

In measuring the quasar image light curves, SCH used two 
nearby stars, *B and *C, as photometric references.
All intensities (i.e.\ for the images as well as for *B) were 
measured with respect to *C, and SCH then subtracted one half 
of the magnitude of *B from the quasar components, for each
observation. The correlation between the quasar and *B may 
result from systematic errors such as the adopted point spread 
function, which varies across the field and so may not be
identical in *B and in the quasar components. In our analysis we 
treat the fluctuations of *B as stochastic noise, with a component 
which is common to *B and to all the quasar components, as
well as an independent source of error. We then let the 
$\chi^2$ minimization decide the size of the error common 
to *B and all the quasar images, separately in each observation, 
but guided by our best estimate of the typical size of the
correlation. Thus, we add *B to our method effectively as
a fourth light curve, but for the underlying model of *B we
take just a constant intensity in time. We also assume
an error component common just to the three quasar images.
Note that such correlated errors are visually apparent from
the data, e.g.\ in Figure 2 of SCH near days 100, 130, and
150, with the clearest example near day 90.

\section{Results for PG1115+080}

Our model for component A consists of L points as described
in \S 2. For B we use the same model, but with an overall
magnitude shift $m_{BA}$ of A with respect to B (accounting for
mean magnification), a time shift $t_{BA}$ due to the time
delay, and an overall variation ratio $\alpha_{BA}$ multiplying the
variation (accounting for differential microlensing as explained in 
\S 2). Similarly, we have $m_{AC}$, $t_{AC}$, and $\alpha_{AC}$,
as well as $m_{BC}$, $t_{BC}$, and $\alpha_{BC}$. When we fit all
the curves simultaneously, we have the constraints
$m_{BC}=m_{BA}+m_{AC}$, $t_{BC}=t_{BA}+t_{AC}$, and
$\alpha_{BC}=\alpha_{BA}\alpha_{AC}$. The only parameters 
which require non-linear $\chi^2$ minimization are $t_{BA}$,
$t_{AC}$, $\alpha_{BA}$, and $\alpha_{AC}$. We minimize with respect
to these parameters using direction set methods (\cite{nr}, 
\S 10.5).

We vary a number of parameters in the model. We try $L=10$,
20 and 30 interpolation points. We try linear, cubic and
cubic spline interpolation. For the measurement errors,
we begin with the uncorrelated errors for A, B and C of 
2.3, 7.5 and 4.0 millimagnitudes (mmag) at $1\sigma$, respectively
(SCH). We also take an uncorrelated error of 1.2 mmag for the
bright star *B.
We then add an error of standard deviation $\sigma_n$ 
common to *B and all QSO components, and also separate
errors of $f_1\sigma_n$ in *B and $f_2\sigma_n$ in all the
QSO components. If we assume $f_1=f_2=1$, then
since *B has a measured variability of 8 mmag
at $1\sigma$, we take $\sigma_n \simeq 6$ mmag. As we show
below, reasonable variations in $\sigma_n$, $f_1$ and
$f_2$ affect the minimum $\chi^2$ value but not appreciably values 
of the derived parameters such as the time delays. On the other
hand, having no correlated errors at all ($\sigma_n=0$) does
change the derived parameters. All the errors are naturally
assumed to be Gaussian in the $\chi^2$ method. 

We can also take Gaussian errors for Monte Carlo trials. On the
other hand, since the method reconstructs a model for
the input signal, it also reconstructs an estimate
of the measurement errors in all the light curves in
each observation. We take these actual errors instead of
Gaussian errors for a second, bootstrap-like Monte Carlo analysis.
Thus we have a set of $36$ observations, and in the Monte
Carlo trials for each day we pick one of the $36$ at random (with
replacement) and add the reconstructed errors on that day (in
A, B, C, and *B) to the matching simulated light curves 
(i.e.\ A, B, C, and *B, respectively). Note that this is not a rigorous 
bootstrap, since after the $\chi^2$ minimization the errors are
no longer independent. However, this procedure should give
us an idea of the actual size and mutual correlations of the
measurement errors, and should be more robust and conservative 
than the Gaussian Monte Carlo trials. For the input signal in
the Monte Carlo trials we take a smooth curve (see Figure 1)
with roughly the same shape as the variation in the actual
data. This way, the input is identical to the real signal
on large time scales, and we assume that any apparent
small scale variability is dominated by measurement
noise, which seems likely.

We define our 'standard' model as having $L=20$, cubic spline
interpolation, $\sigma_n = 6$ mmag and $f_1=f_2=1$. Thus
the assumed covariance matrix of errors for this model in
mmag squared is
\beq
\left(\begin{array}{cccc}
77.3 & 72 & 72 & 36 \\
72 & 128.3 & 72 & 36 \\
72 & 72 & 88.3 & 36 \\
36 & 36 & 36 & 73.3 
\end{array}\right) 
\eeq
with rows from the top in the order A, B, C, and *B. 
Hereafter we use these standard settings except where
otherwise noted.

We begin by fitting each quasar
component separately, always along with *B. Table 1
shows the resulting values of the $\chi^2$ per degree of 
freedom $\bar{\chi}^2$, for various $L$. 
The number of data points is 72 (36 each for a quasar 
component and for *B), and the number of parameters
is $L+1$ (1 for the mean value of *B), which
yields $71-L$ for the number of degrees of freedom (ndof). The
reduced $\chi^2$ is in most cases less than 1, which is
reasonable since we are assuming that a significant portion 
of the error is correlated among all the quasar components,
and if we take each of them separately then we are 
not including this very strong constraint. Table 1 also
suggests that if our error estimates are roughly correct
for A and C, they may be somewhat underestimated for 
component B. Note that the $\bar{\chi}^2$ value
is higher than 1 at $L=10$ only for B, which is the faintest
component and so has the largest errors. This may indicate 
higher non-Gaussianity for the errors in this component, since with 
lower $L$ the fewer parameters cannot effectively
compensate for outlier points. 

Next we fit *B and two quasar components at a time, with the
results displayed in Table 2. There are $36*3=108$ data
points, and $L+4$ parameters (e.g.\ *B mean, $m_{BC}$,
$t_{BC}$ and $\alpha_{BC}$). Also shown are the one sided $68\%$
confidence limits derived from 300 bootstrap Monte Carlo 
trials. Both the bootstrap errors and the parameters
($t_{BC}$ etc.) that we input into the Monte Carlo
trials are in each case (i.e.\ for each pair of
components and for each $L$) set according to the
values fitted from the data in that case. So e.g., for the BC 
pair with $L=20$, a value of $t_{BC}=25.3$ days is used
as input for the Monte Carlo trials which in turn 
determine the error of $^{+2.3}_{-2.0}$ days on $t_{BC}$.
We only show the results of varying $L$ since this leads
to the largest variations (see below). Note that the
$BA$ time delay has the largest fractional uncertainty,
as we expect: $B$ has the largest errors, so the 
uncertainty in $t_{BA}$ is greater than in $t_{AC}$,
and while $C$ has larger errors than $A$, the $BC$
delay is larger and much better resolved than is
$t_{BA}$. For the variation ratios, there is some
variation with $L$, but only $\alpha_{BC}$ is consistent
with 1, while the others disagree with $1$ at the
$4-5\sigma$ level for each $L$. 

Finally we combine all three quasar components,
with results shown in Table 3. Uncertainties,
where shown, are again one sided $68\%$
confidence limits derived from bootstrap 
Monte Carlo trials. All the models
are based on our standard inputs, except for
the changes shown in the first column. In 
rows 4 and 5, $\sigma_n$ is in mmag. In rows
6 and 7, $f_1$ and $f_2$ refer to assumptions
about the correlated errors (see the beginning
of this section). In rows 8 and 9, we use
cubic and linear interpolation, respectively
(see \S 2 above). As above, $r_{ABC}\equiv t_{AC}/t_{BA}$.
We try a large range of model assumptions, most of 
which show very little variation in parameter
values relative to varying $L$. The exception
is the last row, marked 'SCH-like'. Here we
use our method, but with assumptions that correspond 
closely to those of SCH: we assume no correlated
errors, the uncorrelated errors are doubled,
*B is not included in the fitting but
rather we subtract half of *B from the quasar light
curves, and we set the variation ratios to 1.
The 'SCH-like' parameter values are compatible with 
those of SCH to about $1\sigma$.

A $\chi^2$ method, in which errors are assumed to be Gaussian,
can be sensitive to a small number of outliers. We use
two procedures to check for the effects of such 'bad' 
measurements. First, we iteratively remove outliers.
After fitting with our standard model parameters, we
compare each point on every light curve to the reconstructed
signal. We divide the difference by the individual
uncorrelated error for that point, which yields the number of 
standard deviations that each point is off by. We add $20\%$ to
the error for points off by more than $4\sigma$, and add
$10\%$ for points off by $2\sigma-4\sigma$. We then fit
with the adjusted errors, and repeat until convergence.
The resulting values are $t_{BC}=24.0$ days and $r_{ABC}=1.15$, 
well within $1\sigma$ of the results from the standard fit.
As a second check, we try fitting the time delays after removing 
points from the data set. We remove three points at a time,
i.e.\ one point from each light curve, all taken in the same
observation. This yields $36$ sets of data. In two of these data sets
we remove points from the first or last observation, thereby
changing the total time span of the data and thus the model
spacing for a fixed $L$. In order to avoid this, and to be
able to check the effects of outliers independently from the
effects of model spacing, we in fact do not explicitly remove
any points, but rather multiply their individual errors by
$1000$ for a de facto removal. The $36$ fits to these data sets 
yield a mean $t_{BC}=25.1$ days with a standard deviation of
$0.35$ days, a minimum of $23.9$ days and a maximum of $26.1$
days. They also yield a mean $r_{ABC}=1.13$ with a standard 
deviation of $0.03$, a minimum of $1.05$ and a maximum 
of $1.20$. Our two checks suggest that the fitting results are 
{\it not}\/ strongly influenced by outlier points in the data.

For the standard case ($L=20$), there are $4*36=144$
data points, and $L+7$ parameters. In each
case, we can calculate $\alpha_{BC}=\alpha_{BA}\alpha_{AC}$.
E.g.\ for $L=20$ the result is $\alpha_{BC}=1.10^{+.08}_{-.08}$. 
With $L=20$, the $1\sigma$ uncertainties
from Gaussian Monte Carlo trials are, for comparison
with Table 3, $^{+0.9}_{-0.9}$ days for $t_{AC}$, $^{+1.1}
_{-1.1}$ days for $t_{BA}$, $^{+1.4}_{-1.3}$ days for $t_{BC}$, 
$^{+.13}_{-.14}$ for $r_{ABC}$, 
$^{+.07}_{-.06}$ for $\alpha_{AC}$, $^{+.04}_{-.04}$ for $\alpha_{BA}$,
and $^{+.07}_{-.07}$ for $\alpha_{BC}$. The $95\%$ confidence
limits from bootstrap Monte Carlo trials are, also for the
$L=20$ case, $^{+2.0}_{-2.2}$ days for $t_{AC}$, $^{+2.9}
_{-3.2}$ days for $t_{BA}$, $^{+3.3}_{-3.8}$ days for $t_{BC}$, 
$^{+.31}_{-.41}$ for $r_{ABC}$, 
$^{+.16}_{-.20}$ for $\alpha_{AC}$, $^{+.10}_{-.12}$ for $\alpha_{BA}$,
and $^{+.15}_{-.19}$ for $\alpha_{BC}$.
For this standard case, Figure 1 shows the quasar
components A ($\star$), B ($\bullet$), and C ($\circ$)
(including the final point
in each light curve, which was excluded from the fitting).
All have been corrected in each observation only by the 
error component which is common to all the quasar
components and to *B, as reconstructed by the fitting.
Other than this A is shown as observed (except for
a vertical shift to have zero mean), but B and C are
shifted and scaled according to the values of 
$m_{AC}$, $m_{BA}$, $t_{AC}$, $t_{BA}$, $\alpha_{AC}$,
and $\alpha_{BA}$. Also shown are the
reconstructed signal (solid line) and the input
used in the Monte Carlo trials (dotted line).
Figures 2 and 3 show
one dimensional $\chi^2$ plots, as a function
of $t_{BC}$ and $r_{ABC}$, respectively.
In each plot, at every point the parameter shown
is fixed and $\chi^2$ is minimized with respect
to all other parameters. The plots do not show
any strong local minima that could be confused
with the global minimum. The formal $1\sigma$
uncertainties derived from these curves are
$1.0$ day for $t_{BC}$ and $.12$ for $r_{ABC}$.
Figure 4 shows a two dimensional $\chi^2$
plot as a function of $t_{AC}$ and $t_{BA}$,
around the minimum (marked $\times$). The same $\chi^2$
contours are also shown as a function of $t_{BC}$ and
$r_{ABC}$. Note that the result of SCH (marked $\circ$) 
is outside even the outermost contour, which delineates 
the formal $99.99\%$ confidence level. The figure
clearly shows that the main disagreement of the present
results with SCH is in the value of $r_{ABC}$, while
the values of $t_{BC}$ are consistent with each other.

We adopt the $L=20$ results and bootstrap uncertainties,
since they are compatible with the values for the
other $L$ to about $1\sigma$. Thus $t_{BC}=25.0^{+1.5}_{-1.7}$
days, and Tables 2 and 3 indicate that this $6\%$ uncertainty 
(at $68\%$ confidence) is a reasonable estimate. 
Our value used with lens models reduces the induced
$H_0$ only by $5\%$ relative to the SCH value.
On the other hand, $r_{ABC}$ varies fairly strongly
with different assumptions, and is only
weakly constrained at $r_{ABC}=1.13^{+.18}_{-.17}\ .$
This result is close to the values around $1.4$ 
predicted by lens models, but the uncertainty
is too large to be able to decide among
different types of models (\cite{keeton}).
The fitting also recovers $m_{AC}=2.033\pm.005$,
$m_{BA}=-2.534\pm.007$, and $m_{BC}=-.501\pm.008$,
which yield variation-subtracted magnification
ratios for the mean flux. These agree with SCH
and also with the flux ratios of Kristian et al.\ (1993),
since the quasar does not vary much over the time
scale of the time delays. Our extremely accurate
magnification ratios are not useful for lens modelling,
since they are still likely to be greatly altered
by microlensing, as suggested by the variation 
ratios.

Regarding the variation ratios, $\alpha_{BC}=1.10
^{+.08}_{-.08}$ is consistent with 1, while $\alpha_{AC}=
1.39^{+.07}_{-.08}$ is greater than 1 and $\alpha_{BA}=
.79^{+.05}_{-.06}$ is less than 1, both at $4-5\sigma$. 
These values are also roughly consistent with
Table 2. To make perhaps a more direct test of the
significance of this result, we also perform Monte
Carlo tests on input which has variation ratios
equal to 1, but they are allowed to vary in the fitting. 
The result indicates that if the variation ratios were
really 1, we would measure $\alpha_{AC}=1.00^{+.07}_{-.07}$
($^{+.14}_{-.17}$ at $95\%$ confidence) and $\alpha_{BA}=
1.00^{+.10}_{-.10}$ ($^{+.19}_{-.23}$ at $95\%$ confidence), 
so in this sense the
$L=20$ results for $\alpha_{AC}$ and $\alpha_{BA}$ are
significantly different from 1 at $5\sigma$ and 
$2\sigma$, respectively. As discussed in \S 2, this 
may indicate that the quasar optical emission region 
must be of order 1000 AU in linear dimension.

\section{Conclusions}

We have developed a method for determining time delays
among light curves of multiple images of a gravitational
lens. The method constructs a simple model for the
actual source variation, using interpolation between
a number of equally spaced values. It then performs
a combined $\chi^2$ minimization by fitting all of the 
light curves to this model simultaneously, which is
similar to the method of Press et al.\ (1992a, b).
The ability to vary the basic parameters of the
model over a large range lends robustness to our method.
Most of the parameters are linear and so the 
$\chi^2$ minimization is easily done. In addition to
the time delays, the other non-linear parameters are
relative variation ratios, which account for different
fractional variation in flux for different images.
We interpret this physically as evidence for differential
microlensing, i.e.\ a different magnification due to
microlensing for the varying region from the region
giving rise to the mean flux.

Applying our method to the light curves of PG1115+080
observed by SCH, we find a value of $25.0^{+3.3}_{-3.8}$ 
days ($95\%$ confidence) for the 
delay between components B and C, and a ratio $t_{AC}/t_{BA}$
for the two smaller delays of $1.13^{+.18}_{-.17}$
($68\%$ confidence). Unlike SCH,
we include correlated measurement errors as well as
the above mentioned variation ratios in the analysis.
Our result for $t_{BC}$ agrees with SCH, but the
ratio $r_{ABC}$ does not. Our result for $r_{ABC}$ does
agree with lens models, but we find that with the present
data it cannot be derived accurately enough to help
in fitting lens models. For the variation ratios,
we find $\alpha_{AC}=1.39^{+.16}_{-.20}$ and $\alpha_{BA}=.79
^{+.10}_{-.12}$ ($95\%$ confidence),
each indicating differential microlensing at a 
significance of $4-5$ times our estimated $1\sigma$
uncertainties. If confirmed as the data accumulates, 
this would imply that the size of the quasar optical 
emission region is of order 1000 AU,
for microlenses of $\left<M\right>=0.1M_{\sun}$. 
Further data may also allow a determination of the
time variation of the two microlensing magnifications.

\acknowledgements

I thank Paul Schechter for suggesting this project
and for helpful advice, Ed Bertschinger for
encouragement and useful comments, and Doug Richstone
and Ned Wright for valuable comments. This work was 
supported by NASA grant NAG5-2816 and NSF grant
AST-9529154.

\newpage

\begin{deluxetable}{ccccc}
\tablenum{1}
\tablecaption{Single quasar component fitting} 
\tablehead{
\colhead{$L$} & \colhead{ndof} & \colhead{A, $\bar{\chi}^2$}   
& \colhead{B, $\bar{\chi}^2$}    & \colhead{C, $\bar{\chi}^2$} }
\startdata
10 & 61 & 0.66 & 1.15 & 0.66 \nl
20 & 51 & 0.58 & 0.78 & 0.58 \nl
30 & 41 & 0.67 & 0.74 & 0.67 \nl
\enddata
\end{deluxetable}

\begin{deluxetable}{ccccc}
\tablenum{2}
\tablecolumns{5}
\tablecaption{Fitting of quasar components in pairs}
\tablehead{\colhead{$L$} & \colhead{ndof} &
\colhead{$\bar{\chi}^2$} & \colhead{time delay [days]} & \colhead{var ratio} \\
\cline{1-5} \\
\multicolumn{5}{c}{BC components} }
\startdata
10 & 94 & 1.77 & 25.0$^{+1.7}_{-1.8}$ & 1.07$^{+.09}_{-.09}$ \nl
20 & 84 & 1.54 & 25.3$^{+2.3}_{-2.0}$ & 1.09$^{+.09}_{-.09}$ \nl
30 & 74 & 1.03 & 26.7$^{+1.8}_{-1.8}$ & 1.03$^{+.06}_{-.07}$ \nl
\cutinhead{AC components}
10 & 94 & 1.27 & 11.0$^{+0.9}_{-0.9}$ & 1.35$^{+.06}_{-.06}$ \nl
20 & 84 & 1.06 & 11.0$^{+0.9}_{-0.9}$ & 1.52$^{+.09}_{-.08}$ \nl
30 & 74 & 0.98 & 12.3$^{+1.3}_{-1.0}$ & 1.40$^{+.09}_{-.10}$ \nl
\cutinhead{BA components}
10 & 94 & 2.38 & 11.7$^{+2.0}_{-2.0}$ & 0.74$^{+.05}_{-.05}$ \nl
20 & 84 & 1.48 & 6.75$^{+1.7}_{-1.5}$ & 0.61$^{+.05}_{-.04}$ \nl
30 & 74 & 1.06 & 7.22$^{+1.2}_{-1.2}$ & 0.46$^{+.03}_{-.03}$ \nl
\enddata
\end{deluxetable}

\begin{deluxetable}{ccccccccc}
\tablenum{3}
\tablecolumns{9}
\tablecaption{Fitting of the three quasar components}
\tablehead{\colhead{Input\tablenotemark{*}} & 
\colhead{ndof} & \colhead{$\bar{\chi}^2$} & 
\colhead{$t_{AC}$ [days]} & 
\colhead{$t_{BA}$ [days]} & \colhead{$t_{BC}$ [days]} & 
\colhead{$r_{ABC}$} & \colhead{$\alpha_{AC}$} 
& \colhead{$\alpha_{BA}$} \\ \cline{1-9} }
\startdata
$L=10$ & 127 & 1.83 & 12.5$^{+0.9}_{-0.9}$ & 14.0$^{+1.6}_{-1.6}$ &
26.5$^{+1.7}_{-1.7}$ & 0.89$^{+.12}_{-.13}$ & 1.45$^{+.06}_{-.06}$ &
.80$^{+.05}_{-.05}$ \nl
$L=20$ & 117 & 1.60 & 13.3$^{+0.9}_{-1.0}$ & 11.7$^{+1.5}_{-1.6}$
& 25.0$^{+1.5}_{-1.7}$ & 1.13$^{+.18}_{-.17}$ & 1.39$^{+.07}_{-.08}$
& .79$^{+.05}_{-.06}$ \nl
$L=30$ & 107 & 1.38 & 14.9$^{+1.4}_{-1.3}$ & 10.6$^{+1.3}_{-1.3}$
& 25.5$^{+1.8}_{-1.6}$ & 1.41$^{+.27}_{-.25}$ & 1.43$^{+.08}_{-.08}$
& .72$^{+.04}_{-.05}$ \nl
$\sigma_n=4$ & 117 & 2.27 & 13.0 & 11.8 & 24.8
& 1.10 & 1.38 & .80 \nl
$\sigma_n=8$ & 117 & 1.34 & 13.4 & 11.7 & 25.1
& 1.15 & 1.39 & .79  \nl
$f_1=0$ & 117 & 2.01 & 12.8 & 11.4 & 24.2
& 1.12 & 1.34 & .80  \nl
$f_2=0$ & 117 & 2.13 & 12.8 & 11.8 & 24.6
& 1.08 & 1.38 & .81 \nl
cubic & 117 & 1.56 & 13.4 & 11.9 & 25.2
& 1.13 & 1.39 & .80  \nl
linear & 117 & 1.55 & 13.5 & 11.7 & 25.1 
& 1.15 & 1.40 & .79  \nl
SCH-like & 84 & 1.61 & 8.9$^{+1.7}_{-1.6}$ & 11.9$^{+2.8}_{-2.5}$ & 
20.9$^{+2.9}_{-2.6}$ & 0.75$^{+.26}_{-.29}$ & 1 & 1  \nl
\tablenotetext{*}{See text.}
\enddata
\end{deluxetable}

\newpage

\begin{figure}[h]
\vspace*{9 cm}
\caption{Light curves for components A ($\star$),
B ($\bullet$), and C ($\circ$). B and C have been
shifted to match A, and all have been partly corrected
for errors (see text). Also shown are the
reconstructed signal (solid line), the input
used in Monte Carlo trials (dotted line), and the final point
in each light curve, which was excluded from the fitting.}
   \includegraphics{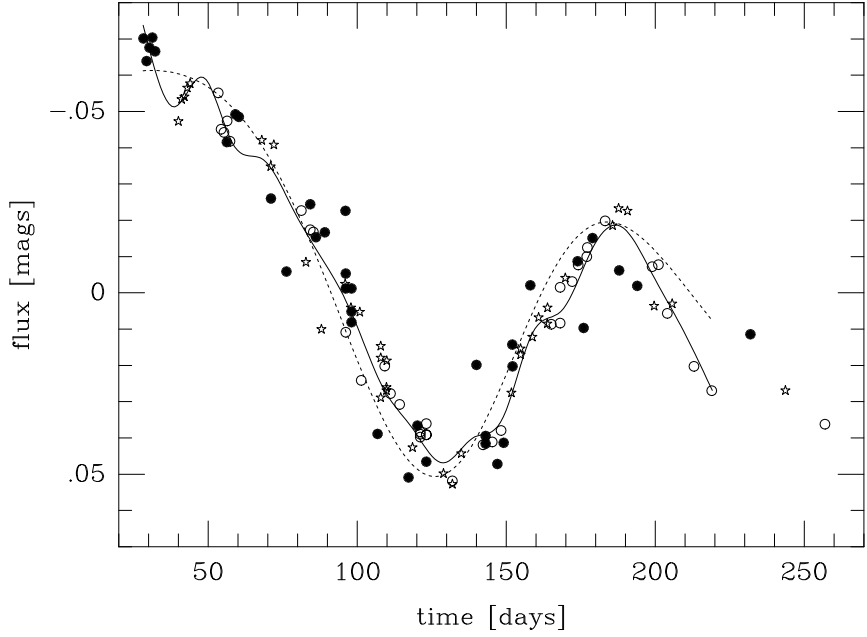}
\end{figure}

\begin{figure}[h]
\vspace*{8.2 cm}
\caption{Value of $\chi^2$ versus assumed delay $t_{BC}$,
relative to the $\chi^2$ value at its global minimum.}
   \includegraphics{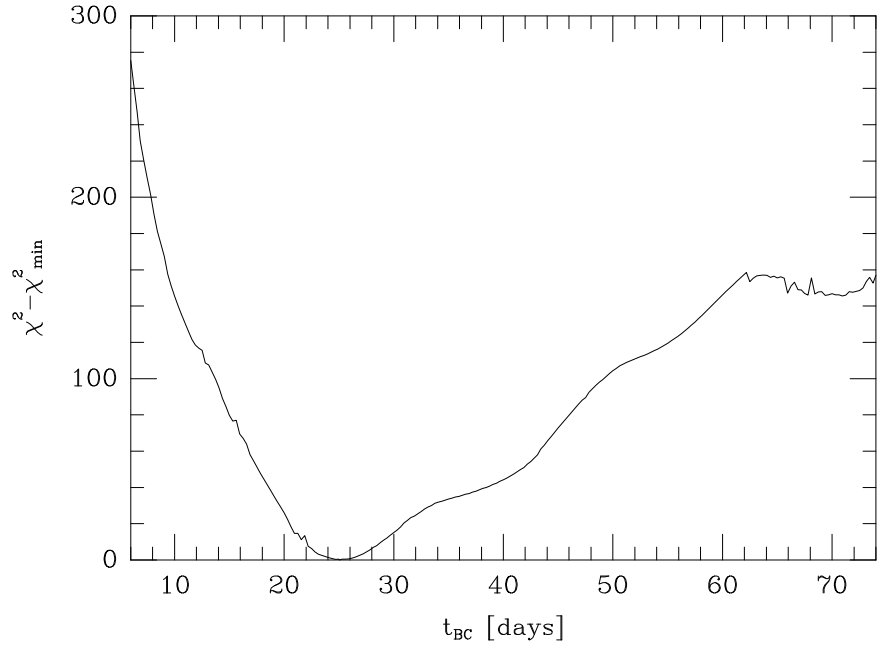}
\end{figure}

\begin{figure}[h]
\vspace*{8.2 cm}
\caption{Value of $\chi^2$ versus assumed ratio $r_{ABC}$,
relative to the $\chi^2$ value at its global minimum.}
   \includegraphics{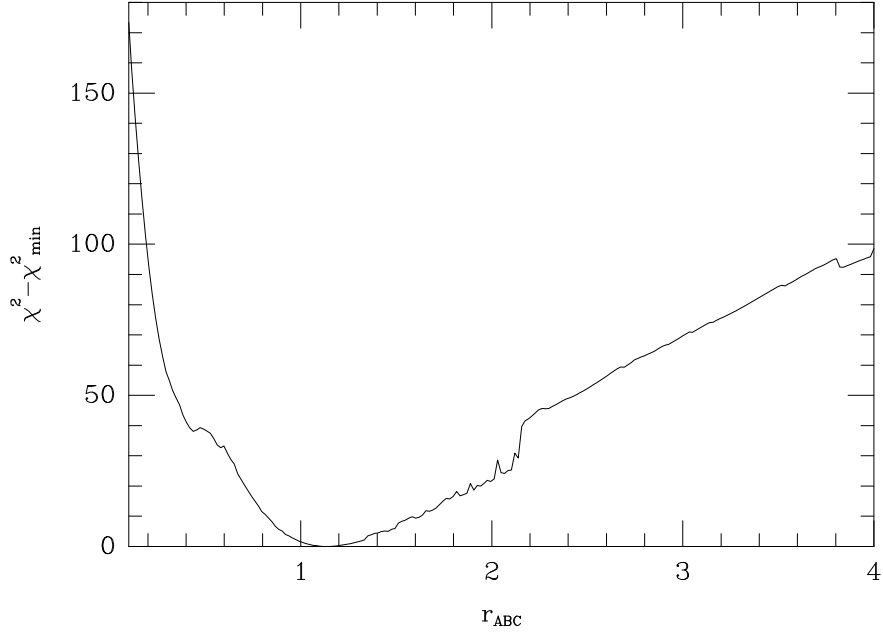}
\end{figure}

\begin{figure}[h]
\vspace*{9 cm}
\caption{$\chi^2$ contours in the $t_{AC}$, $t_{BA}$
plane (left) and in the $t_{BC}$, $r_{ABC}$ plane (right). 
The point marked $\times$ is $\chi^2=187$,
the global mimimum. The contours are drawn at
$\Delta \chi^2=$2.30, 4.61, 6.17, 9.21, 11.8 and
18.4, the $1\sigma$, $90\%$, $2\sigma$, $99\%$,
$3\sigma$ and $99.99\%$ confidence levels for
two parameters. The point marked $\circ$ is the
SCH result.}
   \includegraphics{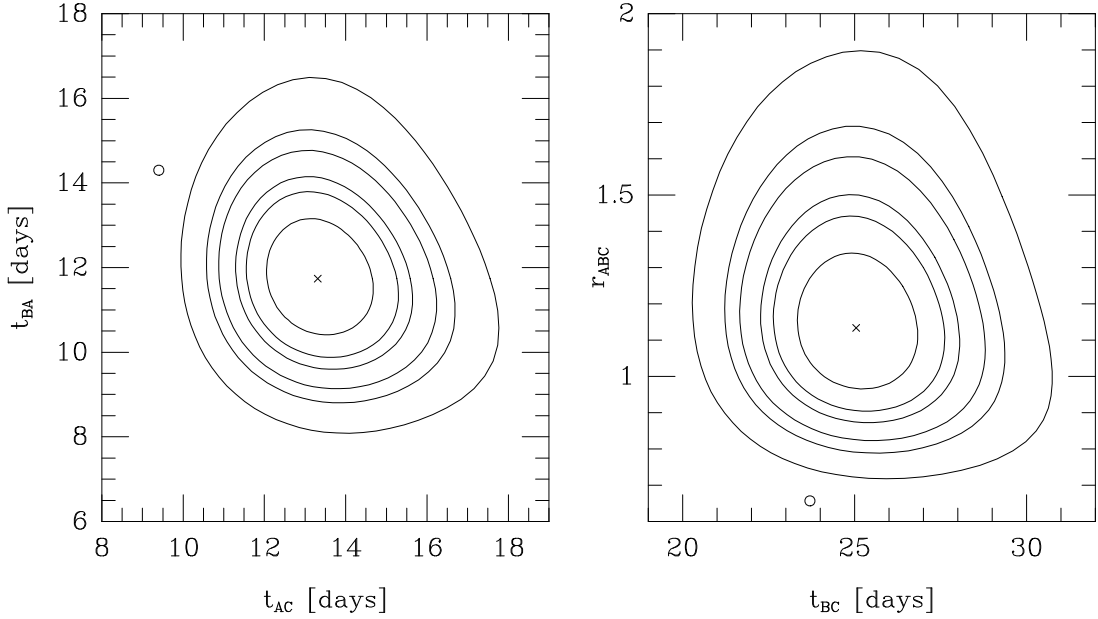}
\end{figure}

\end{document}